\documentstyle[sprocl]{article}

\def\NPB{{\em Nucl. Phys.} B}
\def\PLB{{\em Phys. Lett.}  B}

\def\be{\begin{equation}}
\def\ee{\end{equation}}
\def\bea{\begin{eqnarray}}
\def\eea{\end{eqnarray}}

\newcommand{\tr}[1]{\,{\rm tr}\,#1\,}

\begin{document}

\title{LARGE $N$ QCD
AND $q$-DEFORMED QUANTUM  FIELD THEORIES}

\author{ I.Ya.AREF'EVA }

\address{Steklov Mathematical Institute,\\
Vavilov 42, GSP-1, 117966, Moscow, Russia}

\maketitle\abstracts{
A construction \cite{AVM} of master field describing multicolour QCD is 
presented. The master fields
for large N matrix theories satisfy to standard equations of relativistic field
theory but fields are quantized according 
$q$-deformed commutation relations with $q=0$.
These commutation relations are realized in the Boltzmannian Fock space.
The master field for gauge theory does not take values in a finite-dimensional
Lie algebra, however, there is a non-Abelian gauge symmetry and 
BRST-invariance.}

\section{Introduction}

The large $N$ limit in QCD  where $N$ is the number of colours
enables us to understand qualitatively certain striking
phenomenological features of strong interactions \cite {tH,Ven,Wit}.
To perform an analytical investigation one needs to compute the sum of
all planar diagrams. Summation of planar diagrams has been performed only
in low dimensional space-time \cite {KNN,BIPZ}.

It was suggested  \cite {Wit} that there exists
a master field which dominates the large $N$ limit.
There was an old problem in quantum field theory  how to construct
the master field for the large $N$ limit in QCD.
This problem has been discussed in many works.
The problem has been reconsidered  \cite {GG,Doug}
by using methods of non-commutative 
probability theory \cite {Voi,Ac}.
Gopakumar and Gross \cite {GG} and Douglas \cite {Doug}
have described the master field using a knowledge of  all correlation
functions of a model.

The problem of construction of the master field has been
solved in  \cite {AVM}.
It was shown that the master field
satisfies to standard equations of relativistic field theory but it
is quantized according to  $q$-deformed   relations
\begin{equation}
a(k)a^*(k') -qa^*(k')a(k)=\delta(k-k'),	
	\label{1}
\end{equation}
with $q=0$. Relations ({1}) are 
related with quantum groups \cite{FRT}. About attemps to construct a field 
theory based on (\ref{1})  see \cite{AVI,OWG} and refs therein.
 Operators $a(k),a^*(k')$ for $q=0$
 have a  realization in the
free (Boltzmannian) Fock space. 
Quantum field theory in Boltzmannian Fock space has been considered in
papers \cite {AAKV,AAV,AZsm}.
Some special form of this theory realizes  the master field
 for a subset of  planar diagrams,
for the so called half-planar (HP) diagrams and gives an  analytical summations
of HP diagrams \cite {AAV,AZsm}. HP diagrams are closely related 
with so called parquet approximation \cite{TerM}.

\section{Free Master Field}

We  consider the Minkowski space time.
We use a notation $M^{(in)}$ for the free Minkowski
matrix field. For two-point Wightman functions one has
\begin {equation} 
                                                          \label {1.14}
<0|M^{(in)}_{ij}(x)M^{(in)}_{pq} (y)|0>=\delta_{iq}\delta_{jp}
D^{-}(x-y).
\end   {equation} 

A free scalar Boltzmannian field $\phi^{(in)}(x)$ is
given by
\begin {equation} 
                                                          \label {1.7}
\phi^{(in)}(x)=\frac{1}{(2\pi)^{3/2}}\int \frac{d^{3}k}
{\sqrt{2\omega (k)}}(a^*(k)e^{ikx}+a(k)e^{-ikx}) ,
~\omega (k)= \sqrt{k^{2}+m^{2}}
\end   {equation} 
It is an operator in the Boltzmannian Fock space
with relations
\begin {equation} 
                                                          \label {1.8}
a(k)a^*(k')=\delta^{(3)}(k-k')
\end   {equation} 
and vacuum $|\Omega_0),~ ~ a(k)|\Omega_0)=0$.
An $n$-particle state is created from the vacuum $|\Omega_{0})
=1$ by the usual formula
$
|k_{1},...,k_{n})=a^{*}(k_{1})...a^{*}(k_{n}) |\Omega_{0})
$.
There is no symmetric under permutation of $k_{i}$.

The following
basic relation takes place
\begin {equation} 
                                                          \label {1.B}
\lim _{N \to \infty}
\frac{1}{N^{1+\frac{k}{2}}}<0|\tr(
(M^{(in)}(y_{1}))^{p_{1}}...(M^{(in)}(y_{r}))^{p_{r}})|0>
\end   {equation} 
$$
=(\Omega_{0}|(\phi^{(in)} (y_{1}))^{p_{1}}...(\phi^{(in)} (x_{r})
)^{p_{r}}|\Omega_{0}), ~~~k=p_{1}+...+p_{r}.
$$

\section{Master Field for Interacting Matrix Scalar Field}
To construct the master field for interacting quantum field theory
 \cite {AVM} we have to work in Minkowski
 space-time and use the Yang-Feldman formalism  \cite {BD}.
Let us consider a model of an Hermitian scalar matrix field
$M (x)=(M_{ij} (x)),$ $ i,j=1,...,N$ in the 4-dimensional
Minkowski space-time with the field equations
\begin {equation} 
                                                          \label {1.1}
(\Box + m^{2})M (x)=J(x),~~~J(x)=-\frac{g}{N} M^3 (x),
\end   {equation} 
$g$ is the coupling constant.
One has the Yang-Feldman equation \cite {BD}
\begin {equation} 
                                                          \label {1.3}
M(x)=M^{(in)}(x)+\int D^{ret}(x-y)J(y)dy
\end   {equation} 
where $D^{ret}(x)$ is the retarded Green function for the Klein-Gordon
equation and  $M^{(in)}(x)$ is a  free Bose field. The $U(N)$-invariant
Wightman functions  are 
\begin {equation} 
                                                          \label {1.4}
W(x_{1},...,x_{k})=
\frac{1}{N^{1+\frac{k}{2}}}<0|\tr(M(x_{1})...M(x_{k}))|0>
\end   {equation} 
where   $|0>$ is the Fock vacuum for the free field $M^{(in)}(x)$.

We define the {\bf master} as a solution of the equation
\begin {equation} 
                                                          \label {1.5}
\phi(x)=\phi^{(in)}(x)+\int D^{ret}(x-y)j(y)dy ,~~~
j(x)=-g \phi^3 (x).
\end   {equation} 
The master field  $\phi(x)$ does not have matrix indexes.

 {\it At every order of perturbation theory in the
coupling constant one has the following relation } \cite {AVM}
\begin {equation} 
                                                          \label {1.10}
\lim _{N \to \infty}
\frac{1}{N^{1+\frac{k}{2}}}<0|\tr(M(x_1)...M(x_k))|0>
=(\Omega_{0}|\phi (x_{1})...\phi (x_{k})|\Omega_{0})
\end   {equation} 
{\it where   the field $M (x)$ is defined by  (\ref {1.3})
and  $\phi (x)$ is defined by  (\ref {1.5})}.

\section{Gauge field}
The $U(N)$-invariant
Wightman functions corresponding to $SU(N)$ gauge theory with Lagrangian
\begin {equation} 
                                                          \label {3.1}
L=\tr\{-\frac{1}{4}F_{\mu \nu}^{2}-\frac{1}{2\alpha}
(\partial_{\mu}A_{\mu})^{2}+\bar{c}\partial_{\mu}
\nabla_{\mu}c\}
\end   {equation} 
are defined as
\begin {equation} 
                                                          \label {3.4}
W(x_{1},...,x_{k})=
\frac{1}{N^{1+\frac{k}{2}}}<0|tr(\psi_{i_{1}}(x_{1})...
\psi_{i_{k}}(x_{k}))|0>,
\end   {equation} 
where $\psi_{i}=(A_{\mu},c,\bar {c})$,
$A_{\mu}$ is the gauge field for the $SU(N)$ group,
$c$ and $\bar{c}$ are the Faddeev-Popov ghost fields; $\alpha$
is a gauge fixing parameter. 

According \cite {AVM} the limit of functions (\ref {3.4}) when
$N\to \infty$ can be expressed in terms of the master fields
$B_{\mu}(x)$, $\eta(x)$ and $\bar {\eta}(x)$
corresponding to  $A_{\mu}(x)$,
$c(x)$ and $\bar {c}(x)$ , respectively.
The master fields  satisfy to equations
$$
D_{\mu} {\cal F}_{\mu \nu} +
\frac{1}{\alpha}\partial_{\nu}\partial_{\mu}B_{\mu}+
g\partial_{\nu}\bar{\eta} \eta +g\eta\partial_{\nu}\bar{\eta}=0,
$$
\begin {equation} 
                                                          \label {3.5}
\partial_{\mu}(D_{\mu}\eta)=0,~~~
D_{\mu}(\partial_{\mu}\bar{\eta})=0
\end   {equation} 
where $
{\cal F}_{\mu \nu}=\partial_{\mu}B_{\nu}-\partial_{\nu}B_{\mu}
+g[B_{\mu},B_{\nu}]$, $
D_{\mu}\eta=\partial_{\mu}\eta+g[B_{\mu},\eta].$
These equations have the form of the Yang-Mills equations,
 however, the master fields
$B_{\mu}$, $\eta$, $\bar {\eta}$ do not have  matrix indexes
and they do not take values in a finite dimensional Lie algebra.
The gauge group for the field $B_{\mu}$ is an infinite dimensional group
of unitary operators in the Boltzmannian Fock space.
Eqs. (\ref{3.5})
are invariant under the following BRST transformations
$$\delta B_{\mu}=D_{\mu}\eta \epsilon,~ \delta\eta=\eta^{2}\epsilon,~
\bar{\eta}=-\frac{1}{\alpha}\partial _{\mu}B_{\mu}\epsilon,
$$
where $\epsilon$ is a constant infinitesimal parameter, $\eta\epsilon
+\epsilon\eta=0$, $\bar{\eta}\epsilon +\epsilon\bar{\eta}=0.$

\section *{References}

\end{document}